# LIMIT THEOREMS FOR THE FIBONACCI QUANTUM WALK

**CLEMENT AMPADU**

31 Carrolton Road
Boston, Massachusetts, 02132
USA
e-mail: drampadu@hotmail.com

**Abstract**

We study the discrete-time quantum walk in one-dimension governed by the Fibonacci transformation . We show localization does not occur for the Fibonacci quantum walk by investigating the stationary distribution of the walk, in addition, we obtain the weak limit theorem.

## I. Introduction

Limit theorems for quantum walks has been well studied by many authors [1-19], for example.

Related to quantum physics localization of the quantum has been investigated [20-23], for example.

Let $P(N_t = n)$ be the probability that the quantum walker is at position $n \in Z$ at time $t$, a criterion for localization is the following $\lim\sup_{t\to\infty} P(N_t = 0) > 0$.

In this paper we study the Fibonacci quantum walk as defined by A. Romanelli [24]. The Fibonacci sequence gives rise to rich behaviors in quantum systems. For example in [25] the Fibonacci quantum was shown to produce an unexpected sub-ballistic wave function spreading by the power-law tail of the standard deviation $\sigma(t) \sim t^c$ with $0.5 < c < 1$. This result was also confirmed in [26,27] which studied the quantum walk subjected to noise with a Levy waiting-time distribution [28]. In [29] the Fibonacci quantum walk is shown to produce sub-ballistic behavior for the quantum kicked rotor in resonance, and in [30] tight-binding in electrons.

This paper is organized as follows in Section II we define the Fibonacci quantum walk. In Section III we present the main results with proof. We show non-existence of localization in the Fibonacci quantum walk via Theorem 1, and in Theorem 2 we give the weak limit theorem. Section IV is devoted to an open problem.

**II. Definition of the Fibonacci quantum walk**

The Fibonacci quantum walk is usually generated from a large sequence of two time-step unitary operators $U_1$ and $U_2$ for each time $t$. Given the two initial values of the succession $U_1$ and $U_2$, the sequence is obtained using the rule $U_{k+1} = U_k U_{k-1}$. To obtain the operators $U_1$ and $U_2$, consider the standard quantum walk corresponding to a one-dimensional evolution of a quantum system, in a direction which depends on an additional degree of freedom, the chirality, with two possible states: "left" $|L\rangle$ or "right" $|R\rangle$. Assume that the walker can move freely over a series of interconnected sites labeled by an index $n$. In the quantum walk the motion of the particle is selected by the chirality. At each time step a unitary transformation of the chirality takes place and the walker moves according to its final chirality state. The global Hilbert space of the system is the tensor product $H_s \otimes H_c$ where $H_s$ is the Hilbert space associated to the motion on the line and $H_c$ is the chirality Hilbert space.

Let us call $T_-$ ($T_+$) the operators that move the walker one site to the left (right) on the line in $H_s$ and let $|L\rangle\langle L|$ and $|R\rangle\langle R|$ be the chirality projector operators in $H_c$ and consider the unitary operator $U_i(\theta_i) = (T_- \otimes |L\rangle\langle L| + T_+ \otimes |R\rangle\langle R|) \circ (I \otimes K(\theta_i))$, where $K(\theta_i) = \sigma_z e^{-i\theta_i \sigma_y}$ is a unitary operator acting on $H_c$, $\sigma_y$ and $\sigma_z$ being the standard Pauli matrices, and $I$ is the identity operator in $H_s$, then one step of the quantum walk is given by $|\Psi(t+1)\rangle = U_i(\theta_i)|\Psi(t)\rangle$.

The wave vector $|\Psi(t)\rangle$ is expressed by $|\Psi(t)\rangle = \sum_{n=-\infty}^{\infty} \begin{pmatrix} a_n(t) \\ b_n(t) \end{pmatrix} |n\rangle$, where we have associated the upper (lower) component with the left (right) chirality, the states $|n\rangle$ are eigenstates of the position operator corresponding to the site $n$ on the line. The unitary evolution for $|\Psi(t)\rangle$ corresponding to $|\Psi(t)\rangle = \sum_{n=-\infty}^{\infty} \begin{pmatrix} a_n(t) \\ b_n(t) \end{pmatrix} |n\rangle$ can be written as

$$a_n(t+1) = a_{n+1}(t)\cos\theta + b_{n+1}(t)\sin\theta$$

$$b_n(t+1) = a_{n-1}(t)\sin\theta - b_{n-1}(t)\cos\theta$$

To build the operators $U_1$ and $U_2$ we replace $\theta$ immediately above with $\theta_1$ and $\theta_2$ respectively.

If we create the spatial Fourier transform of the amplitude $(a_n(t), b_n(t))^T$ by multiplying both sides of $a_n(t+1) = a_{n+1}(t)\cos\theta + b_{n+1}(t)\sin\theta$, and $b_n(t+1) = a_{n-1}(t)\sin\theta - b_{n-1}(t)\cos\theta$ by $e^{i\left(\phi - \frac{\pi}{2}\right)n}$,

with $\phi \in [-\pi, \pi]$, and summing over the integer index $n$, we get

$$\begin{pmatrix} F(\phi, t+1) \\ G(\phi, t+1) \end{pmatrix} = M(\phi, \theta) \begin{pmatrix} F(\phi, t) \\ G(\phi, t) \end{pmatrix}, \text{ where } F(\phi, t) = \sum_n e^{i\left(\phi - \frac{\pi}{2}\right)n} a_n(t), \; G(\phi, t) = \sum_n e^{i\left(\phi - \frac{\pi}{2}\right)n} b_n(t),$$

and $M(\phi, \theta) = \begin{pmatrix} ie^{-i\phi}\cos\theta & ie^{-i\phi}\sin\theta \\ ie^{i\phi}\sin\theta & -ie^{i\phi}\cos\theta \end{pmatrix}$. Thus, in the Fourier space the dynamics of the

Fibonacci quantum walk is determined by the unitary matrix $M(\phi, \theta)$. If we call $M_1$ and $M_2$ the matrix $M(\phi, \theta)$ evaluated at $(\phi_1, \theta_1)$ and $(\phi_2, \theta_2)$ respectively, then in the Fourier space the dynamics of the Fibonacci quantum walk is determined as follows: Given the two initial values of the succession $M_1$ and $M_2$ the sequence is obtained using the rule $M_{k+1} = M_k M_{k-1}$.

Notice we can write $M(\phi, \theta) = \begin{pmatrix} e^{-i\phi} & 0 \\ 0 & e^{i\phi} \end{pmatrix} \begin{pmatrix} i\cos\theta & i\sin\theta \\ i\sin\theta & -i\cos\theta \end{pmatrix}$. Let $U = \begin{pmatrix} i\cos\theta & i\sin\theta \\ i\sin\theta & -i\cos\theta \end{pmatrix}$,

then we can write $U = P+Q$, where $P = \begin{pmatrix} i\cos\theta & i\sin\theta \\ 0 & 0 \end{pmatrix}$ and $Q = \begin{pmatrix} 0 & 0 \\ i\sin\theta & -i\cos\theta \end{pmatrix}$. So in

the Fourier space we can write the evolution of the Fibonacci quantum walk as

$|\Psi_{t+1}\rangle = \sum_{n\in Z} |n\rangle \otimes (P|\psi_t(n+1)\rangle + Q|\psi_t(n-1)\rangle)$. Let $\|y\|^2 = \langle y|y\rangle$, then the probability that the

quantum walker $N_t$ is at position $n$ at time $t$ is defined by $P(N_t = n) = \|\psi_t(n)\|^2$. The Fourier transform $|\hat{\Psi}_t(k)\rangle$ of $|\psi_t(n)\rangle$ is defined as $|\hat{\Psi}_t(k)\rangle = \sum_{n\in Z} e^{-ikn}|\psi_t(n)\rangle$. By the inverse Fourier transform we have $|\psi_t(n)\rangle = \int_{-\pi}^{\pi} e^{ikn} |\hat{\Psi}_t(k)\rangle \frac{dk}{2\pi}$. The time evolution of $|\hat{\Psi}_t(k)\rangle$ is

$|\hat{\psi}_{t+1}(k)\rangle = M(k)|\hat{\psi}_t(k)\rangle$, where $M(k) = \begin{pmatrix} ie^{-ik}\cos\theta & ie^{-ik}\sin\theta \\ ie^{ik}\sin\theta & -ie^{ik}\cos\theta \end{pmatrix}$. By induction on $t$, we get

$|\hat{\psi}_t(k)\rangle = M(k)^t |\hat{\Psi}_0(k)\rangle$. In particular, the probability distribution can be written as

$P(N_t = n) = \left\| \int_{-\pi}^{\pi} M(k)^t |\hat{\Psi}_0(k)\rangle e^{ikn} \frac{dk}{2\pi} \right\|^2$. To end this section we should remark that the definition

of the Fibonacci quantum walk here can partly be found in [24].

**III. Main Results**

From $M(k) = \begin{pmatrix} ie^{-ik}\cos\theta & ie^{-ik}\sin\theta \\ ie^{ik}\sin\theta & -ie^{ik}\cos\theta \end{pmatrix}$, it is easily seen that the eigenvalues of

$M(k)$ are given by $\lambda_1(k) = e^{iw(k)}$ and $\lambda_2(k) = e^{-iw(k)}$ where $w(k)$ is determined by

$\sin w(k) = \sqrt{1 - \sin k \cos\theta}$. The normalized eigenvectors corresponding to the eigenvalues

$\lambda_j(k)$, $1 \le j \le 2$ are given by, $V_j(k) = N_j \begin{bmatrix} i(\sin(2k)\cot\theta - \lambda_j e^{-ik} \csc\theta) + \cot\theta \\ 1 \end{bmatrix}$, where $N_j$ is

an appropriate normalization factor. Recall that the degeneracy of the eigenvalues is a necessary condition for localization. However for the Fibonacci quantum walk in this paper, none of the eigenvalues are independent of $k$, so we can start by saying that localization does not occur in the Fibonacci quantum walk. In particular, we expect that our limit theorem does not have a $\delta$ − measure corresponding to localization.

To show localization does not occur, rigorously, let us take the initial state as

$$|\psi_0(n)\rangle = \begin{cases} {}^T[\alpha \ \ \beta], & \text{if } n = 0 \\ {}^T[0 \ \ 0], & \text{if } n \neq 0 \end{cases}, \text{ where } |\alpha|^2 + |\beta|^2 = 1, \text{ and } T \text{ is the}$$

transposed operator. We should note that $|\widehat{\Psi}_0(k)\rangle = |\psi_0(0)\rangle$. Now we evaluate the following limits, $\lim_{t \to \infty} P(N_{2t} = 0)$, $\lim_{t \to \infty} P(N_{2t} = n)$, $\lim_{t \to \infty} P(N_{2t+1} = 0)$, and $\lim_{t \to \infty} P(N_{2t+1} = n)$. We should note that the Fourier transform $|\widehat{\Psi}_0(k)\rangle$ can be expressed by the normalized eigenvectors as

$$|\widehat{\Psi}_0(k)\rangle = \sum_{j=1}^{2} \langle v_j(k)|\widehat{\Psi}_0(k)\rangle |v_j(k)\rangle \text{ which implies that}$$

$$|\widehat{\Psi}_t(k)\rangle = M(k)^t |\widehat{\Psi}_0(k)\rangle = \sum_{j=1}^{2} \lambda_j(k)^t \langle v_j(k)|\widehat{\Psi}_0(k)\rangle |v_j(k)\rangle. \text{ So by the inverse Fourier transform}$$

we get $|\psi_t(n)\rangle = \sum_{j=1}^{2} \int_{-\pi}^{\pi} \lambda_j(k)^t \langle v_j(k)|\widehat{\Psi}_0(k)\rangle |v_j(k)\rangle e^{ikn} \frac{dk}{2\pi}$. If $n \neq 0$, then we recall that

$$\widehat{\Psi}_0(k) = |\psi_0(0)\rangle = {}^T[0 \ \ 0] = \begin{bmatrix} 0 \\ 0 \end{bmatrix}, \text{ and it is easy to check that}$$

$$|\psi_t(n)\rangle = \sum_{j=1}^{2} \int_{-\pi}^{\pi} \lambda_j(k)^t \langle v_j(k)|\widehat{\Psi}_0(k)\rangle |v_j(k)\rangle e^{ikn} \frac{dk}{2\pi} = \begin{bmatrix} 0 \\ 0 \end{bmatrix}. \text{ In particular, if } n \neq 0, \text{ then}$$

$P(N_{2t} = n) = P(N_{2t+1} = n) = 0$, so we have $\lim_{t \to \infty} P(N_{2t} = n) = \lim_{t \to \infty} P(N_{2t+1} = n) = 0$. If $n = 0$,

then we note that $|\psi_{2t(2t+1)}(0)\rangle = \sum_{j=1}^{2} \int_{-\pi}^{\pi} \lambda_j(k)^{2t(2t+1)} \langle v_j(k)|\psi_0(0)\rangle |v_j(k)\rangle \frac{dk}{2\pi}$. In particular we

have to evaluate the following integrals, $\int_{-\pi}^{\pi} \lambda_1(k)^{2t} \langle v_1(k)|\psi_0(0)\rangle |v_1(k)\rangle \frac{dk}{2\pi}$,

$\int_{-\pi}^{\pi} \lambda_2(k)^{2t} \langle v_2(k)|\psi_0(0)\rangle |v_2(k)\rangle \frac{dk}{2\pi}$, $\int_{-\pi}^{\pi} \lambda_1(k)^{2t+1} \langle v_1(k)|\psi_0(0)\rangle |v_1(k)\rangle \frac{dk}{2\pi}$, and

$\int_{-\pi}^{\pi} \lambda_2(k)^{2t} \langle v_2(k)|\psi_0(0)\rangle |v_2(k)\rangle \frac{dk}{2\pi}$. To make the calculation manageable, let us write the

eigenvalues of $M(k)$ for $j=1,2$ as $\lambda_j(k) = e^{iw(k)(-1)^{j+1}}$, then we see that we can write

$$\lambda_j(k)^t \langle v_j(k)|\psi_0(0)\rangle |v_j(k)\rangle = e^{iw(k)(-1)^{j+1}t} \begin{bmatrix} (iN_j \sin(2k)\cot\theta - iN_j e^{-ik}\csc\theta + N_j \cot\theta)(-i\alpha\overline{N}_j \sin(2k)\cot\theta + i\alpha\overline{N}_j e^{ik}\overline{\lambda}_j \csc\theta + \alpha\overline{N}_j \cot\theta + \overline{N}_j \beta) \\ -i\alpha|N_j|^2 \sin(2k)\cot\theta + i\alpha|N_j|^2 e^{ik}\overline{\lambda}_j \csc\theta + \alpha|N_j|^2 \cot\theta + |N_j|^2 \beta \end{bmatrix}$$

So,

$$|\psi_{2t(2t+1)}(0)\rangle = \sum_{j=1}^{2} \int_{-\pi}^{\pi} e^{iw(k)(-1)^{j+1} 2t(2t+1)} \begin{bmatrix} (iN_j \sin(2k)\cot\theta - iN_j e^{-ik}\csc\theta + N_j \cot\theta)(-i\alpha\overline{N}_j \sin(2k)\cot\theta + i\alpha\overline{N}_j e^{ik}\overline{\lambda}_j \csc\theta + \alpha\overline{N}_j \cot\theta + \overline{N}_j \beta) \\ -i\alpha|N_j|^2 \sin(2k)\cot\theta + i\alpha|N_j|^2 e^{ik}\overline{\lambda}_j \csc\theta + \alpha|N_j|^2 \cot\theta + |N_j|^2 \beta \end{bmatrix} \frac{dk}{2\pi}$$

which implies that

$$|\psi_{2t(2t+1)}(0)\rangle = \int_{-\pi}^{\pi} e^{iw(k)2t(2t+1)} \begin{bmatrix} (iN_1 \sin(2k)\cot\theta - iN_1 e^{-ik}\csc\theta + N_1 \cot\theta)(-i\alpha\overline{N}_1 \sin(2k)\cot\theta + i\alpha\overline{N}_1 e^{ik}\overline{\lambda}_1 \csc\theta + \alpha\overline{N}_1 \cot\theta + \overline{N}_1 \beta) \\ -i\alpha|N_1|^2 \sin(2k)\cot\theta + i\alpha|N_1|^2 e^{ik}\overline{\lambda}_1 \csc\theta + \alpha|N_1|^2 \cot\theta + |N_1|^2 \beta \end{bmatrix} \frac{dk}{2\pi}$$

$$+ \int_{-\pi}^{\pi} e^{-iw(k)2t(2t+1)} \begin{bmatrix} (iN_2 \sin(2k)\cot\theta - iN_2 e^{-ik}\csc\theta + N_2 \cot\theta)(-i\alpha\overline{N}_2 \sin(2k)\cot\theta + i\alpha\overline{N}_2 e^{ik}\overline{\lambda}_2 \csc\theta + \alpha\overline{N}_2 \cot\theta + \overline{N}_2 \beta) \\ -i\alpha|N_2|^2 \sin(2k)\cot\theta + i\alpha|N_2|^2 e^{ik}\overline{\lambda}_2 \csc\theta + \alpha|N_2|^2 \cot\theta + |N_2|^2 \beta \end{bmatrix} \frac{dk}{2\pi}$$

It is cleary seen that the integrals in $|\psi_{2t(2t+1)}(0)\rangle$ do not exist, so it certain that $\lim_{t\to\infty} P(N_{2t} = 0)$

and $\lim_{t\to\infty} P(N_{2t} = 0)$ do not exist. Since we have shown that

$\lim_{t\to\infty} P(N_{2t} = 0) = \lim_{t\to\infty} P(N_{2t+1} = 0) = $ Does not exist, and $\lim_{t\to\infty} P(N_{2t} = n) = \lim_{t\to\infty} P(N_{2t+1} = n) = 0$

for any initial state, it follows rigorously that localization does not occur in the Fibonacci quantum walk. In particular, the non-existence of localization is given by the following.

**Theorem 1:** $\lim_{t\to\infty} P(N_{2t} = 0) = \lim_{t\to\infty} P(N_{2t+1} = 0) = Does\ not\ exist$, and

$\lim_{t\to\infty} P(N_{2t} = n) = \lim_{t\to\infty} P(N_{2t+1} = n) = 0$

Now we obtain the limit theorem for the Fibonacci quantum walk. Using the method of Grimmett et.al [31], we see that the $r$ – th moment of $N_t$ is given by

$$E\left((N_t)^r\right) = \sum_{n\in Z} n^r P(N_t = n) = \int_{-\pi}^{\pi} \frac{dk}{2\pi} \langle \Psi_t(k) | (D^r | \Psi_t(k) \rangle) = \int_{-\pi}^{\pi} \sum_{j=1}^{2} (t)_r \lambda_j^{-r}(k) (D\lambda_j(k))^r \left| \langle v_j(k) | \hat{\Psi}_0(k) \rangle \right|^2 + O(t^{r-1})$$

where $D = i\left(\dfrac{d}{dk}\right)$ and $(t)_r = t(t-1)\times\cdots\times(t-r+1)$. Let $h_j(k) = \dfrac{D\lambda_j(k)}{\lambda_j(k)}$, then

$$E\left(\left(\frac{N_t}{t}\right)^r\right) \to \int_\Omega \frac{dk}{2\pi} \sum_{j=1}^{2} h_j(k)^r \left| \langle v_j(k) | \hat{\Psi}_0(k) \rangle \right|^2$$ as $t\to\infty$. Substituting $h_j(k) = x$, we have

$$\lim_{t\to\infty} E\left(\left(\frac{N_t}{t}\right)^r\right) = \int_{-|\gamma_\varepsilon|}^{|\gamma_\varepsilon|} x^r f(x) dx$$, where $f(x) = f_K(x; \gamma_\varepsilon)(c_0 x)$ and $|\gamma_\varepsilon| = |\cos\theta|$. Since $f(x)$ is a

density function, the proof is complete. In particular, we the following.

**Theorem 2:** $\dfrac{N_t}{t} \Rightarrow Z$, where $\Rightarrow$ means weak convergence and $Z$ has the density function

$f(x) = (c_0 x) f_K(x; \cos\theta)$, where $f_K(x; a) = \dfrac{\sqrt{1-|a|^2}}{\pi(1-x^2)\sqrt{|a|^2 - x^2}} I_{(-|a|,|a|)}(x)$,

$I_A(x) = \begin{cases} 1, & if\ x\in A \\ 0, & if\ x\notin A \end{cases}$, and $c_0$ is determined by the initial state of the particle undergoing the

Fibonacci quantum walk in one dimension.

### IV. Open Problem

Consider the quantum walk on the $k$ – dimensional lattice governed by the unitary matrix

$$U = A^{\otimes k}, \text{ where } A = \begin{bmatrix} 0 & 0 & i\cos\theta & i\sin\theta \\ i\sin\theta & -i\cos\theta & 0 & 0 \\ i\cos\theta & i\sin\theta & 0 & 0 \\ 0 & 0 & i\sin\theta & -i\cos\theta \end{bmatrix}$$

. It is an open problem to consider the Fibonacci quantum walk as an $m-$state quantum walk without memory on the $k-$dimensional lattice and obtain the associated limit theorems.